\documentstyle[sprocl,epsf]{article}

\def\Journal#1#2#3#4{{#1} {\bf #2}, #3 (#4)}

\newcommand{\nn}{\nonumber}

\newcommand{\be}{\beta}

\def\PLB{{\em Phys. Lett.}  B}
\def\PRL{\em Phys. Rev. Lett.}
\def\PRD{{\em Phys. Rev.} D}

\def\EPC{{\em Euro. Phys. J.} C}

\def\be{\begin{equation}}
\def\ee{\end{equation}}
\def\bea{\begin{eqnarray}}
\def\eea{\end{eqnarray}}

\begin{document}

\title{New Physics Search via the Higgs Self-Coupling}

\author{Shinya Kanemura$^1$, 
       Shingo, Kiyoura$^{1,2}$, 
      Yasuhiro Okada$^{1,3}$,
         Eibun Senaha$^{1,3}$, 
      \\ C.-P. Yuan$^4$}

\address{$^1$Theory Group, KEK, 1-1 Oho, Tsukuba, Ibaraki 305-0801, Japan\\
         $^2$Department of Radiological Sciences, 
         Ibaraki Prefectural University of Health Sciences,
         Ami, Inashiki, Ibaraki 300-0394, Japan\\
         $^3$Department of Particle and Nuclear Physics, 
         the Graduate University for Advanced Studies, Tsukuba, 
         Ibaraki 305-0081, Japan\\
         $^4$ Department of Physics and Astronomy, Michigan State University, 
         East Lansing, Michigan 48824-1116, USA}

\maketitle\abstracts{
We discuss quantum corrections of new physics to 
the triple coupling of the lightest CP-even Higgs boson in 
the Two Higgs Doublet Model (THDM) and also in the Minimal 
Supersymmetric Standard Model (MSSM). 
In the THDM, quartic contributions of the mass of heavy particles 
in the loop can appear, 
which are not absorbed by renormalization of the Higgs boson mass. 
Such non-decoupling effects on the self-coupling can give 
corrections of ${\cal O}(100\%)$,  
even when all measured Higgs couplings with gauge bosons and 
fermions are consistent with the Standard Model prediction. 
In the MSSM, the loop-corrections decouple in such a scenario.
}
  
\vspace*{-6mm}
\noindent
At Linear Colliders (LC's), the mass generation mechanism will be 
explored by precise measurements of the Higgs couplings with 
fermions and gauge bosons via the Higgs production cross section 
rates and the decay branching ratios. 
The determination of the structure of the 
Higgs potential, however, is only possible by measuring the 
Higgs self-couplings.
The trilinear Higgs boson coupling $\lambda_{hhh}^{}$ can be directly 
measured from the double Higgs boson production processes\cite{eehhx}, 
$e^+e^- \to Z^\ast \to Z hh$ and 
$e^+e^- \to {W^+}^\ast \bar{\nu} {W^-}^\ast \nu \to hh 
\bar{\nu}\nu$, when the Higgs boson ($h$) is light.
At the $e^+e^-$ collider with the energy of 
$500$ GeV ($3$ TeV) and the integrated luminosity of $1$ ab$^{-1}$ 
($5$ ab$^{-1}$), $\lambda_{hhh}^{}$ can be measured by about 20\% (7\%) 
accuracy for the Higgs bosons with the mass of $120$ GeV\cite{battaglia}. 

In this talk, we discuss one-loop corrections to $\lambda_{hhh}$
in the two-Higgs-doublet model (THDM) and in the Minimal 
Supersymmetric Standard Model (MSSM).  
The Higgs sector of the MSSM is a special case of the 
{\it weakly coupled} THDM.
Some models based on the dynamical electroweak symmetry breaking 
also yield the {\it strongly coupled} THDM as their low-energy 
effective theory. 
The $hhh$ coupling in the THDM, in general, differs from that in  
the Standard Model (SM) at tree-level, depending on  
parameters of the Higgs sector\cite{eehhx,hhh_tree}. 
Quantum corrections to the Higgs self-coupling, especially their 
decoupling property, have been studied 
in the MSSM\cite{hhh_hollik1}. 
In the SM, the leading one-loop contribution of the top quarks ($t$) 
to the effective self-coupling is expressed by 
\vspace{-2mm}
\begin{eqnarray}
  \lambda_{hhh}^{eff}(SM) =  \frac{3 m_h^2}{v}
       \left[\, 1 - \frac{N_{c}}{3 \pi^2} \frac{m_t^4}{v^2 m_h^2}
     \left\{ 1 + {\cal O}\left(\frac{m_h^2}{m_t^2},\; 
                              \frac{p_i^2}{m_t^2} \right) \right\} 
\,\right], \label{smhhh}
\label{mt4term_hhh_SM}
\end{eqnarray}

\vspace*{-2mm}
\noindent
where $m_h^{}$ ($m_t^{}$) is the physical mass of $h$ ($t$), 
$N_c$ is the color number of $t$,  
and $p_i^{}$ ($i=1$-$3$) represent momenta of the external lines.
The non-vanishing ${\cal O}(m_t^4)$ term is a striking 
feature of the one-loop correction to the 
self-coupling\footnote{
We note that the top-loop effects on the $hVV$ ($VV=ZZ$, $WW$) couplings 
($g_{hVV}^{}$) and the Yukawa coupling 
can also yield non-decoupling power-like contributions of the top-quark mass, 
but it turns out that the contribution is at most quadratic (not quartic).}.
If a new physics particle has the similar property to the top quarks, 
the new physics effect on $\lambda_{hhh}^{}$ can become large
due to the quartic mass contribution.
We here examine this possibility in the context of the THDM 
and the MSSM.

The Higgs potential of the CP-conserving THDM is given by
\vspace{-3mm}
\begin{eqnarray}
  {V}_{\rm THDM}  =     m_1^2 \left| \Phi_1 \right|^2 
                          + m_2^2 \left| \Phi_2 \right|^2 - 
                              m_3^2 \left( \Phi_1^{\dagger} \Phi_2 
                                + \Phi_2^{\dagger} \Phi_1 \right) 
                                \nn  
                       + \frac{\lambda_1}{2} 
                               \left| \Phi_1 \right|^4 
                             + \frac{\lambda_2}{2} 
                               \left| \Phi_2 \right|^4 \nn 
\end{eqnarray}
\vspace*{-8mm}
\begin{eqnarray}
                          + \lambda_3 \left| \Phi_1 \right|^2 
                                \left| \Phi_2 \right|^2 
                      + \lambda_4 
                               \left| \Phi_1^{\dagger} \Phi_2 \right|^2
                             + \frac{\lambda_5}{2} 
                             \left\{ 
                               \left( \Phi_1^{\dagger} \Phi_2 \right)^2
                            +  \left( \Phi_2^{\dagger} \Phi_1 \right)^2
                             \right\},  \label{pot}
\end{eqnarray}

\vspace*{-3mm}
\noindent
where we imposed a softly-broken discrete symmetry under 
$\Phi_1 \to \Phi_1$ and $\Phi_2 \to - \Phi_2$. 
Two types of the Yukawa interaction are then possible, 
so called Model I and Model II. 
After the diagonalization of the mass matrices, we have 
two CP-even ($h$, $H$), one CP-odd ($A$),  
and a pair of charged ($H^\pm$) Higgs bosons.  
All coupling constants ($\lambda_1$-$\lambda_5$)
are then related to the input parameters $m_h^{}$, $m_H^{}$, $m_A^{}$, 
$m_{H^\pm}^{}$, $\alpha$, $\beta$, $M$ and  $v (\simeq246 {\rm GeV})$, 
where $\alpha$ is the mixing angle between CP-even bosons, 
$\tan\beta=\langle \Phi_2 \rangle/\langle \Phi_1 \rangle$, and 
$M (=m_3/\sqrt{\sin\beta\cos\beta})$ is the soft-breaking 
scale of the discrete symmetry.
The masses of the heavier Higgs bosons 
($H$, $H^\pm$ and $A$) typically have two kinds of origin, as seen by 
\vspace{-3mm}
\begin{eqnarray}
  m_{\Phi}^2 \simeq M^2 + \lambda_i v^2,  \;\;
  (\Phi: H, A, {\rm or}\, H^\pm )
 \label{typ_mass}
\end{eqnarray}

\vspace*{-3mm}
\noindent
which essentially determines the decoupling/non-decoupling 
property of the heavy Higgs bosons\cite{nondec3,nondec_2HDM}.

At LC, $\lambda_{hhh}$ will be measured after precision 
measurements of $m_h$ and $g_{hVV}^2$, so that    
we can use their experimental information to predict 
$\lambda_{hhh}^{eff}$.    
Here, we consider a specific scenario:
(1) Only one light Higgs boson ($h$: CP-even, $m_h=115-160$ GeV) 
    is found. 
(2) Data for the $hVV$ couplings as well as for the Higgs decay 
    branching ratios are in good agreement with the SM prediction. 
This means $\sin^2(\alpha-\beta) \simeq 1$ in the context of the 
THDM\cite{SMlikeTHDM}.
Then, the tree $hhh$ coupling also takes a same form as the SM, ie., 
$\lambda_{hhh}^{tree}(THDM) = 3 m_h^2/v$. 
The leading one-loop contribution is calculated, in this case, 
as\cite{fullpaper} 
\vspace{-4mm}
\begin{eqnarray}
 \lambda_{hhh}^{eff}(THDM) = \frac{3 m_h^2}{v}
      \left\{ 1  
              + \frac{m_{H}^4}{12 \pi^2 m_h^2 v^2} 
                         \left(1 - \frac{M^2}{m_H^2}\right)^3 
              + \frac{m_{A}^4}{12 \pi^2 m_h^2 v^2} 
                         \left(1 - \frac{M^2}{m_A^2}\right)^3 \right.\nn
\end{eqnarray}
\vspace*{-7mm}
\begin{eqnarray}
      \left.        + \frac{m_{H^\pm}^4}{6 \pi^2 m_h^2 v^2} 
                         \left(1 - \frac{M^2}{m_{H^\pm}^2}\right)^3
              - \frac{N_{c} m_t^4}{3 \pi^2 m_h^2 v^2} + 
              {\cal O} \left(\frac{p^2_i m_\Phi^2}{m_h^2 v^2},
                           \;\frac{m_\Phi^2}{v^2},
                           \;\frac{p^2_i m_t^2}{m_h^2 v^2},  
                           \;\frac{m_t^2}{v^2}  \right)
      \right\}. \label{m4THDM}    
\end{eqnarray}
\vspace{-4mm}
The quartic mass terms of the heavier Higgs 
boson masses appear with a supp-

\vspace*{-2mm}
\begin{minipage}[t]{5.7cm}  
\epsfxsize=5.7cm
\begin{center} 
\epsfbox{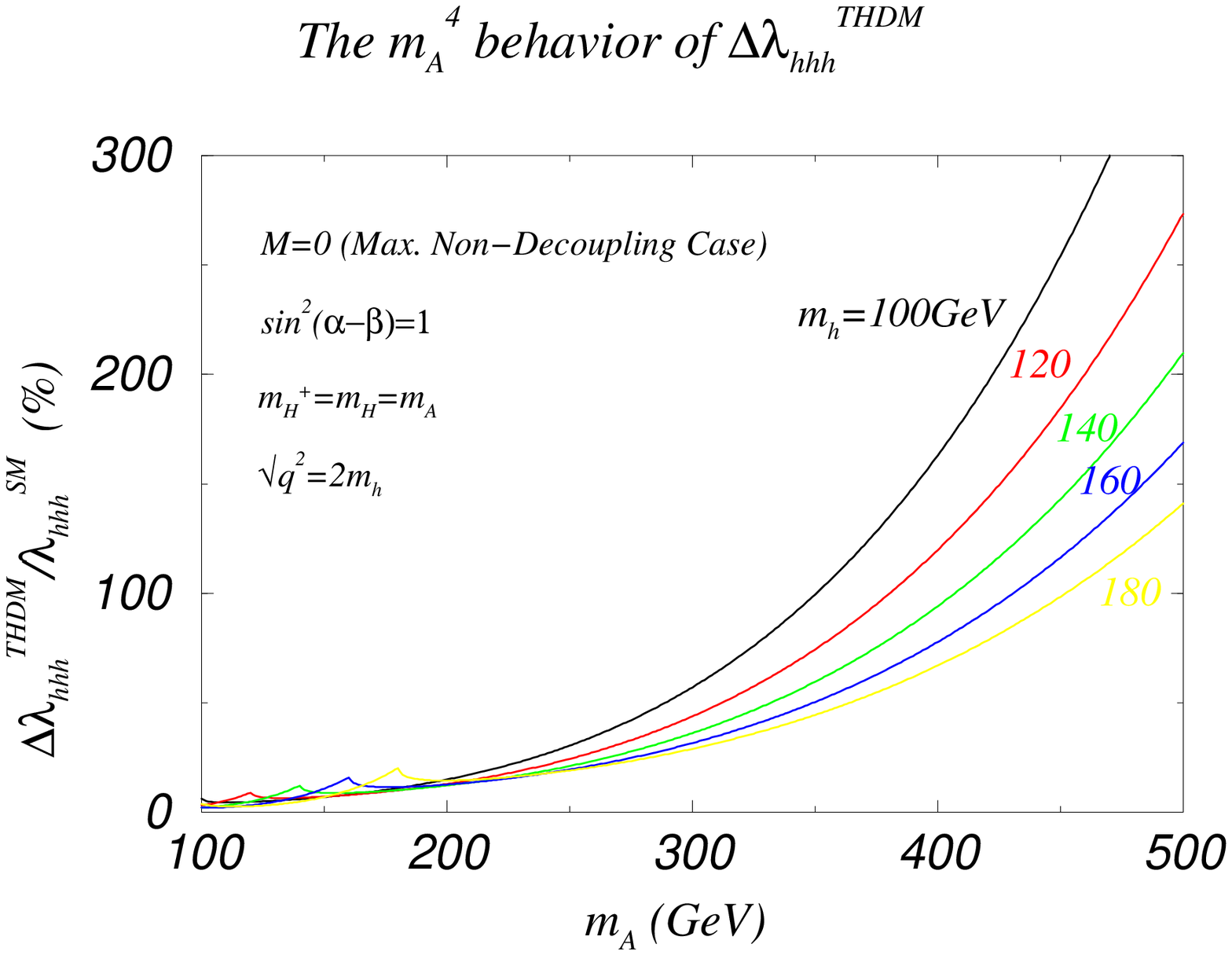}

\vspace*{-3mm}  {\footnotesize \bf Figure 1} 
\end{center} 
\end{minipage}
\begin{minipage}[t]{5.7cm}  
\epsfxsize=5.7cm
\begin{center} 
\epsfbox{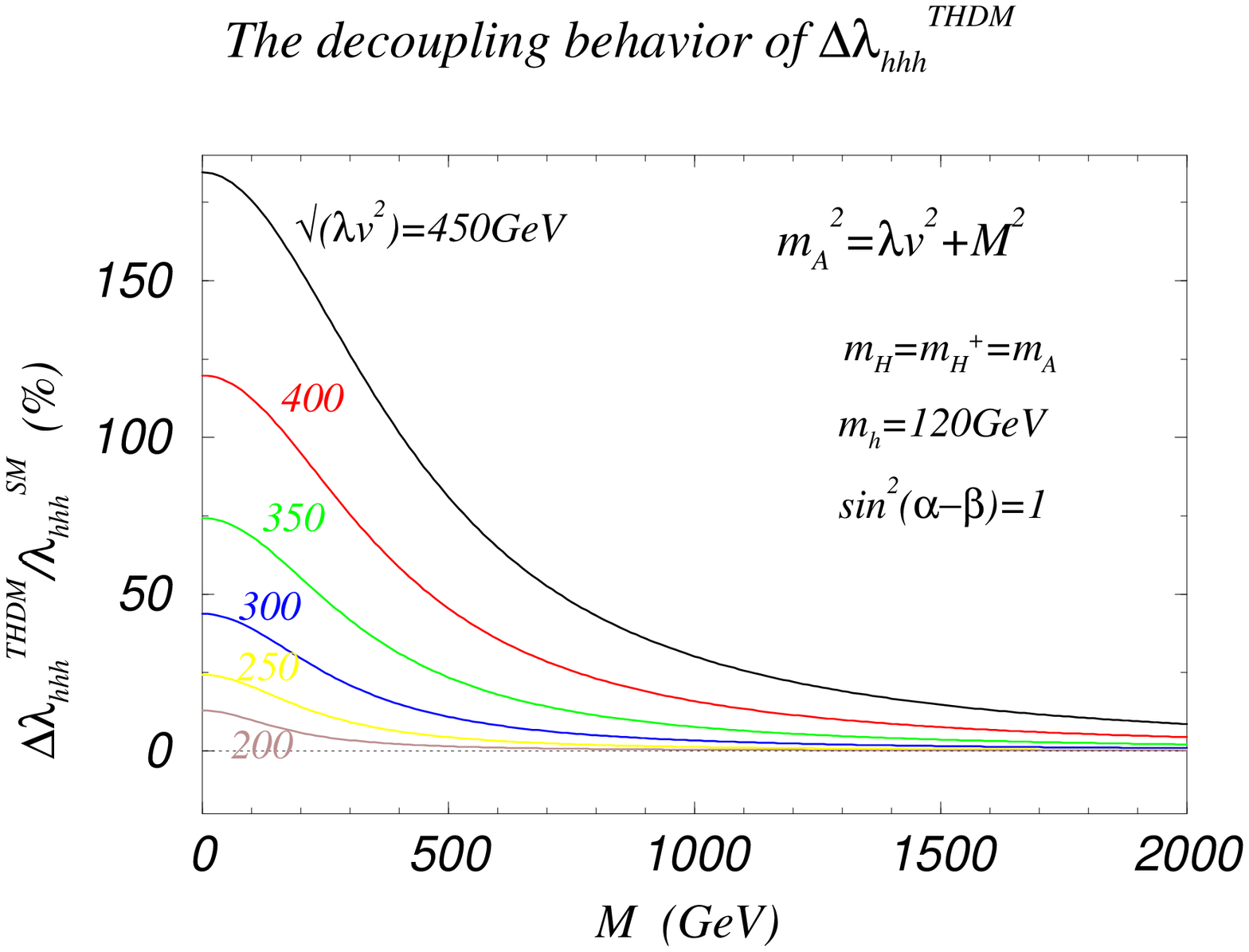}

\vspace*{-3mm}  {\footnotesize \bf Figure 2} 
\end{center} 
\end{minipage}

\noindent
ression factor $(1-M^2/m_{\Phi}^2)^3$, 
reflecting the fact that the THDM has two different parameter region 
regarding the decoupling property. 
\footnote{
For $\sin^2(\alpha-\beta)=1$, the top-loop effect is the same as in the SM, 
being independent of the choice of the Yukawa interaction; 
either Model I or II. 
The difference between Model I and II appears in the region of the 
charged Higgs boson because of the $b\to s\gamma$ results.}

The maximum non-decoupling effects are realized in the limit of $M^2 \to 0$. 
Then, for larger $m_{\Phi}^{}$, greater deviation from the SM prediction is 
obtained as seen in Fig.~1.
Due to the quartic dependences of $m_A^{}$, a large effect
appears for larger values of $m_A^{}$  with smaller $m_h$. 
Since $m_A^{}$ is proportional to $\lambda_i$ in this case, 
large masses are constrained by 
the perturbative unitarity\cite{unitarity} 
($m_A < 580$-$600$ GeV in this case). 
By taking the degeneracy of $m_H^{}=m_A^{}=m_{H^\pm}^{}$, the $\rho$ parameter 
constraint has been taken into account.
The deviation from the SM prediction becomes 
about 30\% (100\%) for $m_A^{}=300$ (400) GeV. 
The decoupling behavior of the heavier Higgs boson contribution
is shown in Fig.~2 as a function of $M$ with $m_A^2=\lambda v^2 + M^2$. 
For a given $M$, the heavy Higgs boson contribution reduces 
rapidly for larger $M \gg \lambda v^2$, though it can still be 
at a few tens of percent level when $M=1$ TeV.  %
The momentum dependence in the effective self-coupling  
$\lambda_{hhh}^{eff}(q^2)$ for $h^\ast \to hh$, 
where $q^\mu$ is the momentum of $h^\ast$,  
is shown for $M=0$ in Fig.~3.

In the MSSM, $\lambda_i v^2$ in Eq.~(\ref{typ_mass}) is constrained to be 
$\lambda_i v^2 \simeq {\cal O}(m_W^2)$, so that the corrections are small 
and the typical decoupling behavior is observed for the loop 
contributions of heavier Higgs bosons\cite{hhh_hollik1}.  
The leading stop-loop effect can be expressed 
in the limit of $m_A \to \infty$ for 
$M_Q^{}=M_U^{}$ ($\equiv M_S^{}$) by 
\vspace{-4mm}
\begin{eqnarray}
  \left.
 \frac{\lambda_{hhh}^{eff}(MSSM)}{\lambda_{hhh}^{eff}(SM)}
  \right|_{p_i=0} \!\!\!\! = \!
1 + \frac{N_{c} m_t^4}{3 \pi^2 v^2 m_h^2} \frac{m_t^2}{M^2_S} 
                     \left(
                      1
                     - \frac{3}{2} 
                       \frac{X_t^2}{M^2_S} 
                     + \frac{1}{2} 
                       \frac{X_t^4}{M^4_S} 
                     - \frac{1}{20} 
                       \frac{X_t^6}{M^6_S} \right), 
\end{eqnarray}

\vspace*{-3mm}
\noindent
where $M_Q^{}$ and $M_U^{}$ represent the soft-SUSY-breaking masses 
of left- ($M_Q^{}$) and right- ($M_U^{}$) handed stops, 
and $m_t |X_t|$ is the off-diagonal element of the stop 
mass matrix. 
The ${\cal O}(m_t^4)$ terms appear with the suppression factor 
$(m_t^2/M_S^2)$, 
so that the effects decouple in the SM-like 
scenario ($M_S^{} \to \infty$).  In Fig.~4, the

\begin{minipage}[t]{5.5cm}  
\epsfxsize=5.5cm
\begin{center} 
\epsfbox{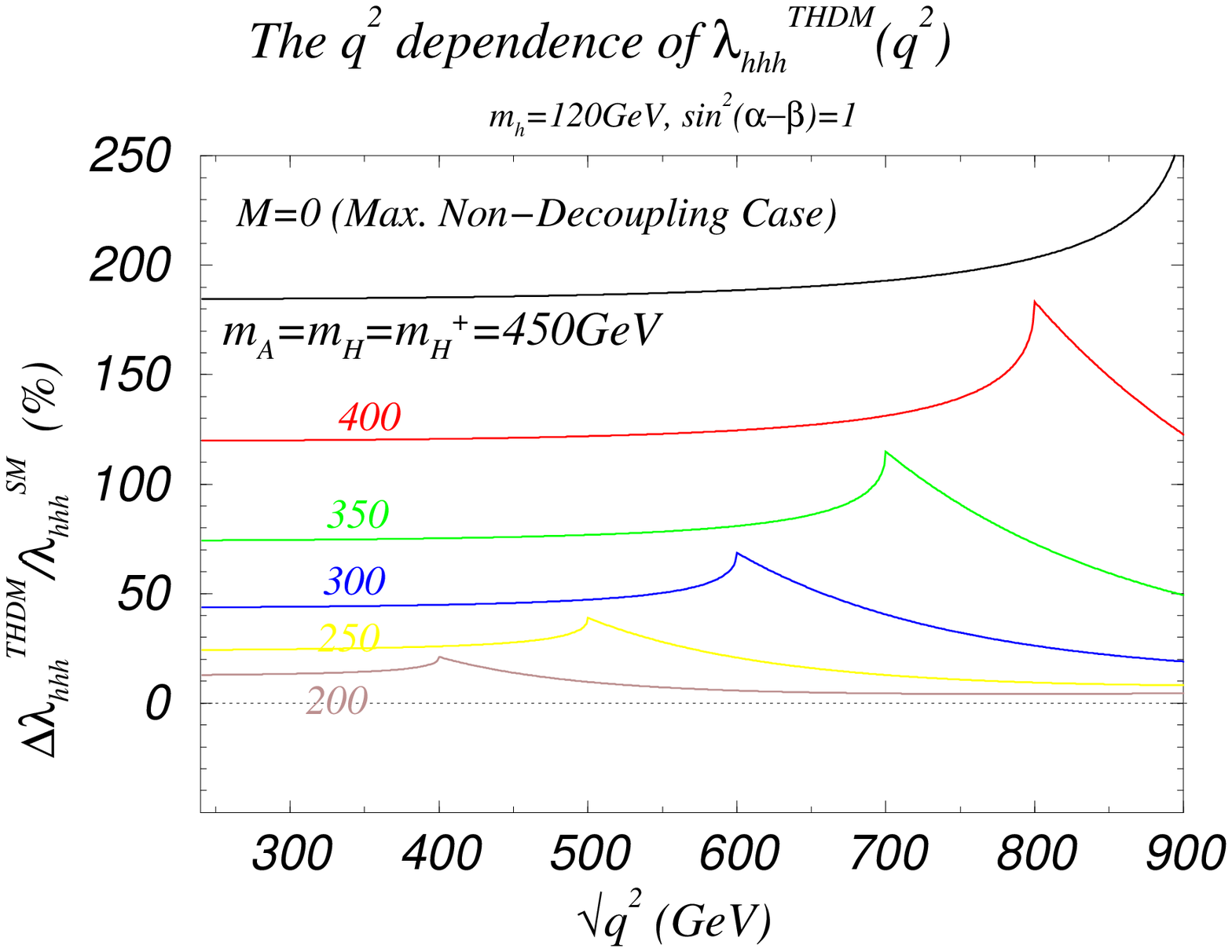}

\vspace*{-3mm}  {\footnotesize\bf Figure 3} 
\end{center} 
\end{minipage}
\begin{minipage}[t]{5.5cm}  
\epsfxsize=5.5cm
\begin{center} 
\epsfbox{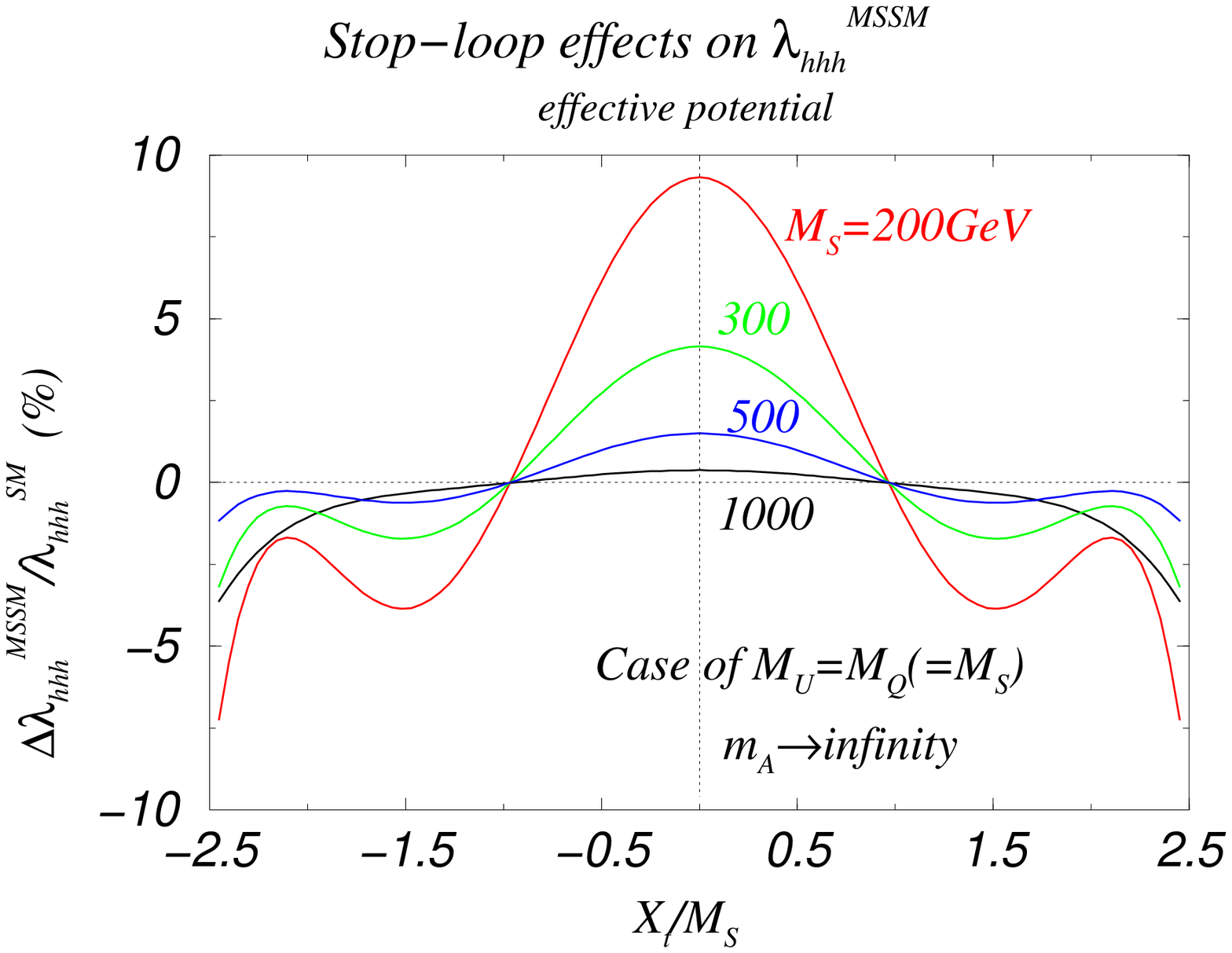}

\vspace*{-3mm}  {\footnotesize\bf Figure 4} 
\end{center} 
\end{minipage}

\noindent
leading stop effect is plotted as 
a function of $X_t^{}/M_S^{}$ for several values of $M_S^{}$. 
  
In summary, the quantum correction can be ${\cal O}(100\%)$ 
in the general THDM due to the quartic mass terms of heavier 
Higgs bosons, even when all the data before the measurement 
of $hhh$ coupling are almost SM like. 
Such large effects may be detected at LC's.
When the Higgs bosons are {\it weakly coupled}, the loop effects 
of the heavier Higgs bosons are suppressed in the large mass limit.
In the MSSM, the Higgs sector is similar to a weakly coupled THDM, so 
that the loop effects of the heavier Higgs bosons are small. 
For the light stop scenarios, the stop-loop contribution can be large, 
and its correction can exceed 5 \%.

\end{document}